\documentclass[a4paper,twoside]{article}

\usepackage{epsfig}
\usepackage{subcaption}
\usepackage{calc}
\usepackage{amssymb}
\usepackage{amstext}
\usepackage{amsmath}
\usepackage{amsthm}
\usepackage{multicol}
\usepackage{pslatex}
\usepackage{apalike}
\usepackage[bottom]{footmisc}
\usepackage{todonotes}
\usepackage{SCITEPRESS}     

\begin{document}

\title{Evaluation Scheme to Analyze Keystroke Dynamics Methods}

\author{\authorname{Anastasia Dimaratos\sup{1} and Daniela Pöhn\sup{2}}
\affiliation{\sup{1}Ludwig Maximilian University of Munich, Munich, Germany}
\affiliation{\sup{2}Universität der Bundeswehr München, Neubiberg, Germany}
\email{anastasia.dimaratos@campus.lmu.de, daniela.poehn@unibw.de}
}


\abstract{
Password authentication is a weak point for security as passwords are easily stolen and a user may ignore the security by using a simple password. Therefore, services increasingly demand a second factor. While this may enhance security, it comes with a lower level of usability and another factor to be forgotten. A smartphone is an important device in daily life. With the growing number of sensors and features in a smartphone, keystroke dynamics may provide an easy-to-use method. In this paper, we introduce requirements for biometric authentication and keystroke dynamics. This results in an evaluation scheme, which is applied to three selected approaches. Based on the comparison, keystroke dynamics and the evaluation scheme are discussed. The obtained results indicate that keystroke dynamics can be used as another authentication method but can be bypassed by stronger adversaries. For further research, a common data set would improve the comparability.
}

\onecolumn \maketitle \normalsize \setcounter{footnote}{0} \vfill

\section{\uppercase{Introduction}}
\label{sec:introduction}

Authentication is required to access several services. It can generally be grouped into the categories of knowledge (e.\,g., a password), possession (e.g., a hardware token), and biometrics (e.\,g., fingerprint or keystroke dynamics). Although prone to several attacks such as guessing, dictionary, and credential stuffing attacks, 90\% of the 1,000 most-visited websites apply passwords~\cite{10.1007/978-3-642-17197-0_13}. With the uptake of mobile computing, password-based authentication is even more problematic due to the nature and reduced level of security of the devices. Given the increasing amount of accounts and related sensitive information, such as credit card information, confidential emails, and personal photos, stored on mobile devices, there is an increasing demand for stronger authentication methods.

To address the emerging threats to password-based mobile authentication, biometric authentication systems are increasingly used for mobile devices. With the growing number of sensors and diverse sets of user-driven features, modern mobile devices offer a platform for capturing and analyzing biometrics for authentication. Biometric authentication identifies users by relying on their physiological or behavioral characteristics. Physiological biometrics, such as fingerprints, iris, finger vein patterns, and face geometry, verify users, who belong to a large population. On the other hand, behavioral biometrics uses the unique pattern and signature of a user to access a service or device. Therefore, the system must identify the illegitimate user even if the correct username and password are entered. 
Keystroke dynamics refers to the unique patterns of rhythm and timing-based features that are created when a user types on a touchscreen and, therefore, uses the sensors and other sets of mobile devices~\cite{MONROSE2000351}. Personal constitution (e.g., stressed, depressed, or drunk), as well as other external factors, can influence the way of typing. In addition, different adversaries have to be considered~\cite{10.1145/3477601}. In order to discuss keystroke dynamics and the conditional influence on the method, an evaluation scheme is required.

Our main contribution is three-fold: i) We assess the effectiveness of keystroke dynamics by establishing a generic evaluation scheme. ii) We use this evaluation scheme to analyze three exemplary keystroke dynamics approaches. iii) Based on this comparison, we discuss the potential of keystroke dynamics.

First, we explain keystroke dynamics, before we summarize related work in Section~\ref{sec:sota}. In Section~\ref{sec:methodology}, we introduce the requirements for biometric authentication and keystroke dynamic systems. These evaluation criteria are applied to compare three approaches in Section~\ref{sec:evaluation}. Based on the comparison, we discuss keystroke dynamics in Section~\ref{sec:discussion}. Last but not least, we draw conclusions and discuss future directions.

\section{\uppercase{Keystroke Dynamics}}
\label{sec:background}

\begin{figure}[htb]
	\centering
	\includegraphics[width=0.4\textwidth]{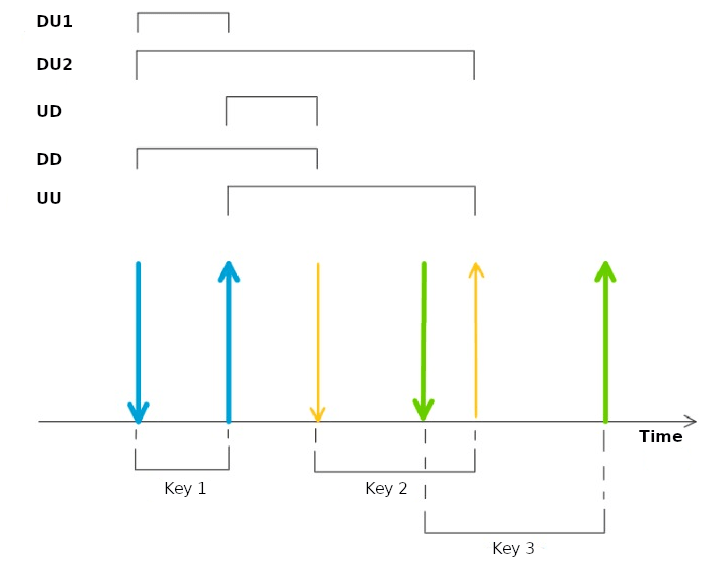}\\ 
	\caption{Features according to \cite{Pisani2013}}
	\label{fig:diagram}
\end{figure}

In this section, the background of keystroke dynamics as a behavioral biometrics authentication method is given. There are two ways to implement it: somebody has to enter a password or a random text either with or without a username. If neither password nor username is needed, then the method is the only factor; In other cases, it is multi-factor authentication. 
In order to use keystroke dynamics, different characteristics (time-dependent and time-independent) are measured. Time-dependent features are different during a time span or between a Down- or an Up-Press. Figure~\ref{fig:diagram} illustrates the most relevant time-dependent features Down-Up-Press (DU1), the Up-Down-Press (UD), the Down-nextUp-Press (DU2), the Down-Down-Press (DD), and the Up-Up-Press (UU). Time-independent features exist more often for modern touchscreen devices than for computers due to the number of sensors. These include the exact x/y-position of the keystroke, the area size of the touch, and the displacement of the finger or pen during a keystroke. For new devices, the position of the mobile phone can be measured, such as the angle or the rotation of the device. Other time-independent features, which are possible for conventional keyboards, include frequent typos and their improvements.

Like most biometric authentication systems, keystroke dynamic systems work in two phases: training and classification. Both phases go through the following four steps, where steps three and four can be implemented by machine learning or statistics:
 
\begin{enumerate}
	\item Data collection;
	\item Preprocessing the data; 
	\item Feature selection and extraction; 
	\item Classification.
\end{enumerate}

For the training of the system, a template for each user is created on the basis of one or more data sets and the resulting features. In the classification stage, the same happens with a new data set. Then, based on probabilities, the system decides whether a user is legitimate or, if no username input is assumed, who the user is. Because of the dynamic nature of the typing behavior, some studies implemented the sixth step of relearning for the keystroke dynamics system (KDS). The data is constantly updated and dynamically tailored to a user through adaptation mechanisms.

\section{\uppercase{Related Work}}
\label{sec:sota}

While several approaches try to improve keystroke dynamics, only a few evaluate them with the same rates and datasets. Zamsheva et al.~\cite{10.1145/3415048.3416118} compare two databases with personal data based on their BehavioSenseapproach. In order to evaluate the accuracy, they use false acceptance rate (FAR), false rejection rate (FRR) as well as global, individual, and average equal error rate (EER). These rates are also applied by other approaches, e.\,g., \cite{5593258,alghamdi2015dynamic,Lee2018,bio}.
Pisani and Lorena systematically analyze keystroke dynamics approaches. The authors use DU1, DU2, UD, DD, and UU classifiers. In addition, FAR, FRR, EER, accuracy rate, and integrated error are applied to evaluate the performance. Last but not least, the authors state benchmarking datasets. A similar survey was conducted by Teh et al.~\cite{Teh2013}, focusing on FAR, FRR, and EER for static and dynamic approaches. As both surveys were published in 2013, newer approaches are not included.
Newer publications either propose an approach (e.\,g., \cite{8967112}) or concentrate on specific issues, including a comparison of different models~\cite{9182115}, machine learning classifiers~\cite{electronics10141622}, and emotion recognition~\cite{9155004}.
In addition, Shekhawat and Bhatt~\cite{9004312} analyze the ERR in multiple use cases and different classification algorithms without converting the validation parameters for comparison.

\section{\uppercase{Methodology}}
\label{sec:methodology}

In this section, we provide the methodology to evaluate keystroke dynamics.

\subsection{Biometric Authentication}
\label{sec:reqbiometric}

We conducted a literature study by searching for ``requirements'' + ``biometric'' at major publishers. This includes \cite{article,7330160,7781835,9693400,10.1145/3437880.3460410,10.1145/3052973.3055160}. We then extracted the following main requirements:

\begin{enumerate}
	\item Universality: As many people as possible should be able to use it.
	\item Uniqueness: A characteristic may not describe two or more people at the same time.
	\item Circumvention: Difficult-to-imitate individual characteristics must be selected by which a human being can be uniquely identified to ensure reliable differentiation.
	\item Permanence: A feature must be selected that is hardly subject to change.
	\item Measurability: The user's biometrics must be measurable and quantifiable.
	\item Security: The system must be robust, i.\,e. non-intrusive.
	\item User-friendliness: The system must be easy to use for the user.
	\item Acceptability: The users must give their consent to the use of the procedure.
	\item Economic feasibility: The cost of a biometric authentication system must be reasonable.
\end{enumerate}

Furthermore, additional requirements can be set for the use of smartphones. A biometric authentication system should be device-independent regardless of the sensor technology. Authentication must also be possible when switching devices, e.\,g., between the work mobile phone and the private smartphone.

\subsection{Authentication Process}
\label{sec:reqauthn}

The authentication process consists of two phases. In the training phase, the characteristics of a person are stored in a database after data collection, pre-processing, and extraction of the characteristics. The actual authentication of the person takes place in the test phase, where the extracted features are compared with the data stored in the database.

\paragraph{Data Collection and Relearning:} Different sensors are used to extract biometric data. The choice of keyboard type (numeric keypad or virtual keyboard) and sensors are factors for data collection. A number field offers fewer possibilities than a virtual keyboard on a smartphone. The accuracy increases with the number of possible entries and thus with the choice of password or input. It can be concluded that:

\begin{itemize}
	\item The type of input and password has an influence on security.
	\item The accuracy of the classification increases with the number of recordings.
	\item Different environmental factors as well as psychological and physical changes result in different behaviors.
	\item  Since behavior changes over the course of a lifetime and frequent repetition, the system should dynamically adjust.
\end{itemize}

\paragraph{Pre-Processing:} Data is not always of the same quality, which is why it should be pre-processed to allow for better comparison. If data is hardly or poorly pre-processed, the following steps will be more difficult, longer, and less accurate.

\paragraph{Feature Selection, Extraction, and Classification:} By extracting and selecting features, the characteristics of a person are defined. During classification, a new input is compared with the profiles and a decision is made regarding legitimacy. The error rates (Section~\ref{sec:Ev}), which represent security and usability, are dependent on these process steps and are good indicators for the usability of a KDS.

\subsection{Evaluation Criteria}
\label{sec:Ev}

The effectiveness of a biometric authentication method (i.\,e., distinction between legitimate and illegitimate users) is determined by various error rates. 
In the following, these evaluation criteria are explained.

\subsubsection{Measurable Units of Binary Classification}

For binary classification, the following measurable units are given. To measure the data, the inputs are divided into true positives (TP), false positives (FP), false negatives (FN), and true negatives (TP).

\paragraph{Accuracy (AC)}
Accuracy is 
defined as the number of correct predictions divided by the total number of predictions as follows:
\begin{footnotesize}
\begin{equation}
	accuracy = \frac{TP+TN}{TP+TN+FP+FN}
\end{equation}
\end{footnotesize}
\vspace{-0.6cm}

\paragraph{Precision (P)}
The precision answers the question of what proportion of positive identifications are correct. This can be determined by FP and TP:
\begin{footnotesize}
\begin{equation}
	precision = \frac{TP}{TP+FP}
\end{equation}
\end{footnotesize}
\vspace{-0.6cm}

\paragraph{Recall (R)}
The recall describes the proportion of actual positives correctly identified, determined by FN and TP:
\begin{footnotesize}
\begin{equation}
	recall = \frac{TP}{TP+FN}
\end{equation}
\end{footnotesize}
\vspace{-0.6cm}

\paragraph{F1-score}
The F1-score verifies the accuracy of a binary classification model for a data set:
\begin{footnotesize}
\begin{equation}
	F1 = 2 * \frac{precision*recall}{TP+FN}
\end{equation}
\end{footnotesize}
\vspace{-0.6cm}

\subsubsection{Comparison Rates}
\begin{figure}[htb]
	\centering
	\includegraphics[width=0.45\textwidth]{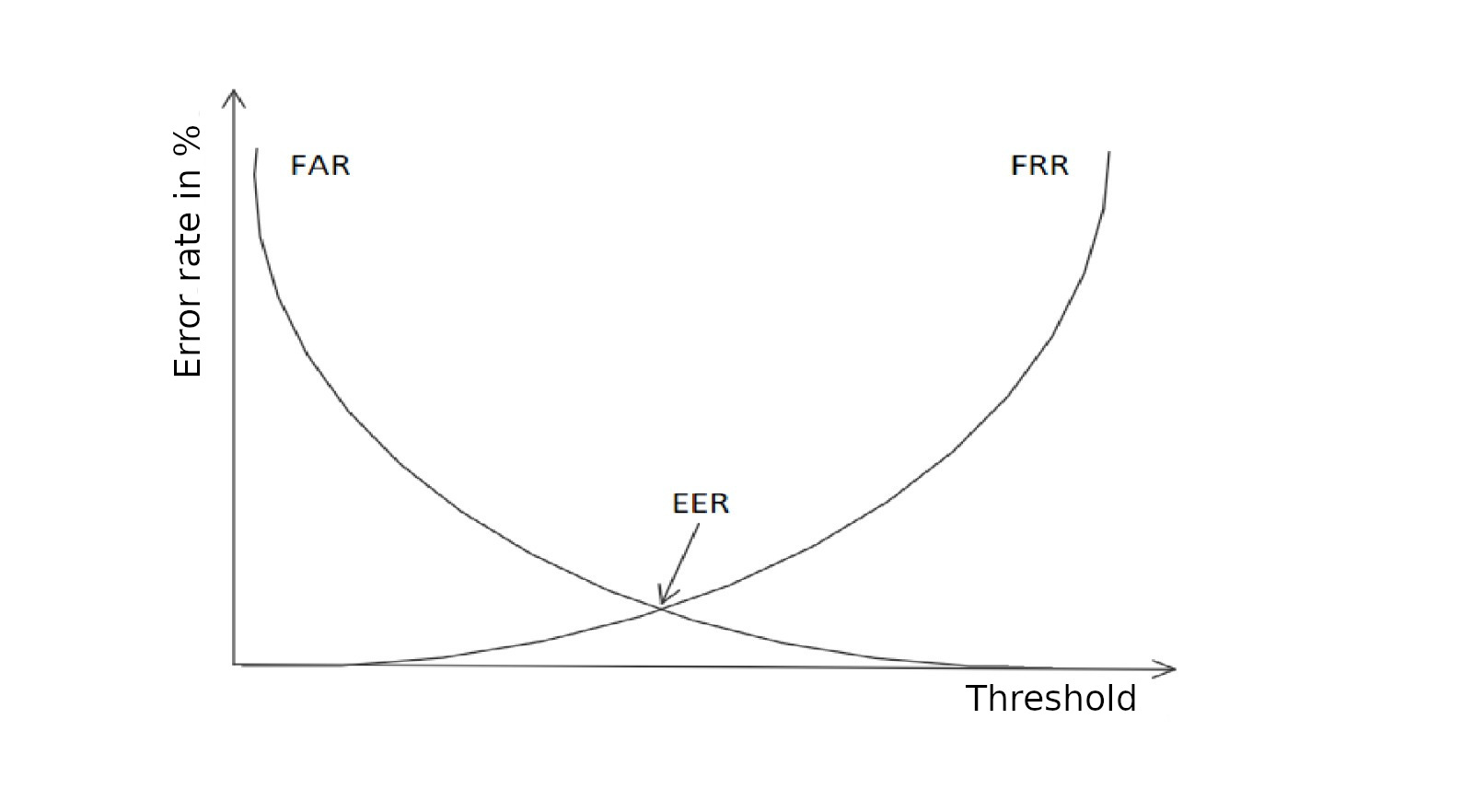}\\ 
	\caption{Correlation of error rates}
	\label{fig:EER}
\end{figure}

Different error rates are applied~\cite{5593258,Pisani2013,alghamdi2015dynamic,Lee2018}. Figure~\ref{fig:EER} shows the correlation of the error rates explained below.

\paragraph{Error}
An error describes the incorrect assignment of an example by the classifier.

\paragraph{False Rejection Rate (FRR)}
FRR defines the legitimate users who are incorrectly rejected. This rate is also called the False Nonmatch Rate (FNR), False Alarm Rate, False Positive Rate (FRP), or Type 1 Error. FRR depends on the threshold set, which indicates the degree of correspondence of the new input to the reference data from which a user is recognized as legitimate. 
According to the European Standard for Access Control (EN-50133), an authentication method should not reach a value for FRR above $1\%$ to ensure security under European law.
\begin{footnotesize}
\begin{equation}
	FRR = \frac{Total\ number\ incorrectly\ rejected\ users}{Total\ number\ login\ attempts\ legimitate\ users}
	\label{FormelFRR}
\end{equation}
\end{footnotesize}
\vspace{-0.6cm}

\paragraph{False Acceptance Rate (FAR)}
FAR defines the percentage between the false granting of access by fraudsters and the total number of fraudsters accessing the system. 
Often it is also called False Match Rate (FRM) or Type 2 error. This rate depends on the threshold set for the acceptance of a user. 
According to the European Standard for Access Control (EN-50133), an authentication method should not reach a value above $0.001\%$ to ensure security.
\begin{footnotesize}
\begin{equation}
	FAR = \frac{Total\ number\ incorrectly\ accepted\ user}{Total\ number\ login\ attempts\ illlegimitate\ users}
	\label{FormelFAR}
\end{equation}
\end{footnotesize}
\vspace{-0.6cm}

\paragraph{Equal Error Rate}
In order to enable overall accuracy, EER or also called Crossover Error Rate (CER) is used. EER is the intersection between FRR and FAR (Figure \ref{fig:ErklärEER}).

\paragraph{Failure to Enrol Rate (FER)}
If a feature of biometrics, such as typing behavior, cannot be used
, then FER is a metric for this. 
\begin{footnotesize}
\begin{equation}
	FER = \frac{Number\ of\ persons\ with\ failed\ data\ recording}{Total\ number\ of\ potential\ users}
	\label{BerFER}
\end{equation}
\end{footnotesize}
\vspace{-0.6cm}

\paragraph{Failure to Acquire Rate (FTA)}
FTA is a rate for erroneous data recording and missing data generation in comparison with the reference data.
\begin{footnotesize}
\begin{equation}
	FTA = \frac{Counts\ of\ failed\ data\ recordings\ after\ traing}{Total\ number\ of\ potential\ users}
	\label{BerFTA}
\end{equation}
\end{footnotesize}
\vspace{-0.6cm}

\subsubsection{Significance of the Comparative Rates}
Due to the dependence of the comparison rates on the data collected, statements should be viewed with caution when comparing these from different studies. FAR, FRR, and EER depend not only on the data collected but also on the threshold set. For an adequate statement, the threshold value is necessary to know. 
An attacker starts with the probability of success $p$ on each attempt. If $FAR = p$ for an input attempt corresponds to $0.01$ and $FRR = q = 0.02$, then FAR and FRR change as follows on the second attempt:
\begin{footnotesize}
\begin{equation}
	\begin{split}
		FAR_2 = p + (1-p) \cdot p \\\
		FAR_2 = 0.01 + (1-0.01) \cdot 0.01 = 0.0199
	\end{split}
\end{equation}
\begin{equation}
	\begin{split}
		FRR_2 = q \cdot q \\\
		FRR_2 = 0.02\cdot0.02 = 0.0004
	\end{split}
\end{equation}
\end{footnotesize}
FAR, i.\,e., the ease of use, increases, and FRR, i.\,e., the safety, decreases with each further permitted attempt after a failure. 
For a better comparison of the evaluation criteria, the interrelationships between them are shown below.

\paragraph{Accuracy and EER}
Depending on the EER, the accuracy can be calculated as follows:
\begin{footnotesize}
\begin{equation}
	Accuracy = 1 - EER
	\label{formulaaccuracy2}
\end{equation}
\end{footnotesize}
\vspace{-0.6cm}
\paragraph{EER, FRR and FAR}
The EER is calculated by the intersection of FAR and FRR:
\begin{footnotesize}
\begin{equation}
	EER = 0.5(FAR + FRR)
	\label{FormulaEER2}
\end{equation}
\end{footnotesize}
Because of the conditions set for FAR and FRR, the maximum value for EER is as follows:
\begin{footnotesize}
\begin{equation}
	\begin{split}
		max(EER) = 0.5(0.01 + 0.001) \\\
			\Rightarrow max(EER) = 0.5 \cdot 0.011\\
			\Rightarrow max(EER) = 0.0055\\
			\Rightarrow max(EER) = 0.55\%
		\end{split}
		\label{FormulaEER3}
	\end{equation}
\end{footnotesize}´
\vspace{-0.6cm}

\subsubsection{Various Biometric Features}

To compare different biometric features, the following differentiation possibilities can be used.

\paragraph{Evaluation Criteria of a Biometric Feature}
\begin{enumerate}
	\item Comfort: Includes user-friendliness.
	\item Accuracy: Determines error rates in practical use.
	\item Availability: Describes the potential user group.
	\item Costs: Incurred through data collection.
\end{enumerate}

\paragraph{Possibility of Imitating a Feature}
If an attacker wants to mimic a person's biometric feature, this can vary in difficulty depending on the feature. 
Features can be categorized as follows:
\begin{enumerate}
	\item Open: Feature can be observed without aids.
	\item Slightly hidden: An observer can find the feature.
	\item Covert: Feature can be detected by a specific detector.
	\item Severely hidden: The feature is not directly observable but defined by the results of functions.
\end{enumerate}

\paragraph{Effort of the Attack}
An attack on a system can be subject to different difficulties:
\begin{enumerate}
	\item Low effort: This exists if an accidental and unintentional penetration of the system is possible without prior knowledge, with simple means, and without a major expenditure of time.
	\item Medium effort: If the attack lasts several hours to days and the knowledge of general access information is assumed, it is a medium effort on the part of the attacker.
	\item High effort: If good specialist knowledge or higher 
is required and an attack lasts several weeks, requires good opportunities, and operating resources, then an attack has a high effort.
\end{enumerate}

\section{\uppercase{Evaluation}}
\label{sec:evaluation}


\subsection{Support Vector Machine and Radial Basis Kernel}
\label{sec:concept1}

\cite{krishnamoorthy2018identification} propose a concept with a Support Vector Machine (SVM) and Radial Basis Kernel (RBF). For data collection, they use their own app for Android devices. The 94 participants had to type in the password \texttt{.tie5Roanl} 30 times. Pre-processing consists of five steps: 1) import the data and remove duplicates; 2) exclude data below a threshold; 3) delete geolocation information; 4) search action types and improved errors; 5) generate features for each keystroke and action type. 13 feature types are extracted, resulting in 155 different features for 77 users. For classification, linear SVM is analyzed. In order to improve the accuracy, the authors apply RBF and compare it with SVM only. 

\subsection{Median Vector Proximity}
\label{sec:concept2}

\cite{al2016statistical} apply dynamic static methods with median vector proximity. The 17 participants had to type in a message five times. The recorded data by the Android app is normalized and min-max scaled. Next, seven features are extracted. For classification, two vectors are established. In addition, different thresholds are tested to find the best proportion between FAR and FRR.

\subsection{Distance Vector Classification}
\label{sec:concept3}

\cite{alghamdi2015dynamic} use Distance Vector Classification (DVC) alone and in combination with other methods. The authors programmed an Android app, where 22 participants had to type in the password \texttt{766420} 100 times. For preprocessing, the authors evaluate min-max scaling, standard scale, euclidean distance, and manhattan distance. They decide on the manhattan distance and standard scale. For each participant, 120 features are extracted. The authors evaluate different methods for the classification: DVC and One-Class Support Vector Machine (OCSVM). With DVC, better results are gained.

\subsection{Analysis of the Approaches}

\paragraph{Data Collection}

The different approaches are summarized in Table~\ref{tab:datacollection}. The number of participants and inputs is comparably small, with the second concept having the lowest. 
Also, the keyboards differ: keyboards chosen by the developers up to virtual keyboards and number fields.

\begin{table}[h]
\centering
\caption{Data collection of the chosen approaches}
	\begin{tabular}{l|l|l|l|c}
		\textbf{Con.} & \textbf{Input}             & \textbf{People}   & \textbf{Input} & \textbf{Features} \\ \hline
		\ref{sec:concept1}                & .tie5Roanl & 94 & 30           & 155                                                                                                            \\
		\ref{sec:concept2}                & choosable                  & 17 & 5  & 31/33                    \\ 
		\ref{sec:concept3}                 & 766420                   & 22 & 100        & 120                                                                                                           
	\end{tabular}
\label{tab:datacollection}
\end{table}

\paragraph{Pre-Processing}

As shown in Table~\ref{tab:preprocessing}, the methods for pre-processing differ. 

\begin{table}[h]
	\centering
	\caption{Pre-processing of the approaches}
	\begin{tabular}{l|p{5cm}}
		\textbf{Concept} & \textbf{Methods for pre-processing}                                                                  \\ \hline
		\ref{sec:concept1}    & Removal of duplicates, threshold, and three further steps \\ \hline
		\ref{sec:concept2}    & Normalization, min-max scale                                                                                                      \\ \hline
		\ref{sec:concept3}    & Manhatten distance, standard scale                                                                                                                                                                       
	\end{tabular}
\label{tab:preprocessing}
\end{table}

\paragraph{Selection and Extraction of Features}

Table~\ref{tab:features} provides an overview of the feature types and amounts. Concept 2 (x: length of input) requires fewer features than concepts 1 and 3. 
All concepts apply time-dependent and independent features. Concept 3 additionally takes device-specific features into account.

\begin{table}[h]
	\centering
	\caption{Feature type and amount for each concept}
	\begin{tabular}{l|l|l|l}
		\textbf{Feature type}                                                              & \textbf{\ref{sec:concept1}}    & \textbf{\ref{sec:concept2}}   & \textbf{\ref{sec:concept3}}                                          \\ 
		\hline
		Down-Up                                                                                      & \#16  & \#x   & \#0                                         \\ 
		\hline
		Up-Down                                                                                      & \#15  & \#x-1 & \#6                                         \\ 
		\hline
		Down-Down                                                                                    & \#15 & \#x-1 & \#6                                         \\ 
		\hline
		Up-Up                                                                                        & \#15  & \#0   & \#6                                         \\ 
		\hline
		Down-Up (2-graph)                                                                            & \#15  & \#0   & \#6                                        \\ 
		\hline
		Pressure                                                                                        & \#16  & \#x   & \#0                                        \\ 
		\hline
		Size                                                                                        & \#16  & \#x   & \#6           \\ 
		\hline
		X-Y P                                                                                        & \#16  & \#0   & \#0                                         \\ 
		\hline
		X-Y C                                                                                        & \#16  & \#0   & \#0                                         \\ 
		\hline
		Avg. time & \#5   & \#0   & \#0                                         \\ 
		\hline
		Avg. pressure                      & \#1  & \#1   & \#0                                        \\ 
		\hline
		Avg. size                                                                           & \#1   & \#1   & \#0                                        \\ 
		\hline
		device-specific                       & \#8   & \#0   & \#18  \\ 
		\hline
		D\_n                                                                                         & \#0   & \#0   &\#6                                        \\ 
		\hline
		XyDn                                                                                         & \#0  & \#0   & \#12                                        \\ 
		\hline
		XyUp                                                                                         & \#0   & \#0   & \#12                                        \\ 
		\hline
		TOTAL                                                                                        & \#155 & \#5x  & \#120
	\end{tabular}
	\label{tab:features}
\end{table}

\paragraph{Classification}

While concept 1 uses SVM with RBF, both other concepts apply distance vector classification and evaluate different combinations (Table~\ref{tab:classification}).

\begin{table}[h]
	\centering
	\caption{Applied classification}
	\begin{tabular}{l|p{5cm}}
		\textbf{Concept}       &  \textbf{Classification}    \\ \hline
		\ref{sec:concept1} & SVM with RBF \\
		\ref{sec:concept2} & Distance vector with two vectors \\
		\ref{sec:concept3} & Distance vector with median \\
	\end{tabular}
\label{tab:classification}
\end{table}

\paragraph{Comparison by Evaluation Criteria}

The classification with SVM gains better results if used together with RBF. The inclusion of further features improves the results of the classification. In addition, sex and gender may have an impact on rotation, pressure, size, and acceleration. 
In order to compare the three selected approaches, we apply our evaluation criteria as shown in Table~\ref{tab:comparison}. By converting the results of concept 1 to EER (0.026), we notice that the approach would fulfill the requirements for an authentication method. 

\begin{table}[h]
	\centering
	\caption{Comparison by evaluation criteria}
	\label{tab:comparison}
	\begin{tabular}{l|l|p{3cm}}
			\textbf{Concept} & \textbf{Features}         & \textbf{Evaluation criteria}               \\ \hline
			\ref{sec:concept1} & 36                      & Accuracy $= 0.9740$\newline F1-score $= 0.9701$                          \\ \hline
			\ref{sec:concept2}    & 31/33 &\o EER  = 0.1219 \newline EER von $\sigma$ = 0.1337             \\ \hline
			\ref{sec:concept3}     & 20                      & \o EER = $0.0789$                                                                                 
		\end{tabular}
\end{table}

\section{\uppercase{Discussion}}
\label{sec:discussion}

In this section, we discuss the evaluation metrics and keystroke dynamics.

\subsection{Evaluation Metrics}

Based on the evaluation, we consider the requirements proposed in Section~\ref{sec:Ev}. The first concept applies different criteria (accuracy and F1 score) than the others (FRR and EER). After converting the values according to formula~\ref{formulaaccuracy2}, the accuracy corresponds to an EER of 0.026. 
Even though the values are comparable, the data acquisition, the data sets, and the overall setup of the experiments are not. This means that one or several templates for experiments and pre-defined data sets would help to evaluate several approaches in a structured way. 
According to the results, authentication with a given input performs better than with a free input. The use of the SVM seems to achieve better results than the statistical methods. However, since the values depend on many factors within the study and some of these are not disclosed, a comparison is difficult. Here, again, an open access policy would improve a comparison.

\subsection{Keystroke Dynamics}


\subsubsection{Requirements for Biometric Systems}

The established requirements for biometric systems are partly fulfilled by keystroke dynamics:

\begin{enumerate}
	\item \textbf{Universality:} 
The following counter-example proves that generality is not fully given. A blind or hands-free person, e.\,g., can operate a smartphone using the language assistant. However, if the typing behavior becomes a condition of the authentication, the system may not able to use it because the voice assistant automatically enters the data.
	\item \textbf{Uniqueness:} 
By the fluctuation of the typing behavior of a person and determining by similarities, there is consequently the possibility for a characteristic to describe several people. By combining it with other characteristics, such as the pressure level, the probability to differentiate person $A$ from person $B$ increases.
	\item \textbf{Circumvention:}
Due to the fluctuation of the typing behavior, a broader range is possible, which is why a KDS works with similarity probabilities. The more accurate and
longer several people are analyzed in their behavior, the more likely it is that precise statements about different behavior patterns of a person can be made.
	\item \textbf{Permanence:} 
KDS is a behavior-based characteristic, which changes depending on the psychological and/or physical constitution. For example, a person in an emergency situation wants access to their smartphone to dial the emergency number. It is likely that the tremors, which can cause excitement, deny access.
	\item \textbf{Measurability:} 
The ability to measure and detect is given with a KDS since the characteristics are determined and recorded by sensors, while other features, e.\,g., pressure or coordinates, can be calculated. 
	\item\textbf{Security:} 
This is not easy to measure for a KDS as several factors play a role. A poorly set up system has more security gaps than a securely designed KDS and vice versa.
	\item \textbf{User-friendliness:}
Since for the use of the KDS, no additional hardware is required and users do not have to learn anything new, regardless of whether it is a computer keyboard or virtual keyboard on the smartphone, the KDS can be seen as easy to use.
	\item  \textbf{Acceptability:} 
If a user does not accept the analysis of their typing behavior and is convinced that entering the password by using a password manager is safer, the method of keystroke dynamics fails. For keystroke dynamics, no further means are required. However, the procedure is little known to the public, so it is difficult to make a statement about it.
	\item \textbf{Economic feasibility:} 
In order to implement authentication using KDS, there are a few costs: no specific hardware, but software. Extensive user training is also not necessary.
\end{enumerate}

Not all requirements are fulfilled. 
A biometric authentication system should be independent of the sensor technology applied in the smartphone. This may vary dependent on the KDS -- and partly even on the scenario (e.\,g., broken screen). Therefore, a new learning phase per device/scenario may be needed.

\subsubsection{Risks}

A biometric authentication system consists of different steps, namely the data acquisition by means of sensors, the pre-processing of the data, the extraction, the classification, and the storage of the reference data in a database or the comparison with the reference data resp. the decision. Inside and between all these steps, the system is vulnerable. In the case of keystroke dynamics, the vulnerable system is typically part of the smartphone.

The accuracy is medium accurate with a maximum of 97\% according to the selected papers. Due to the changeable nature of the typing behavior, a permanent analysis may be more precise. The typing behavior cannot be simply imitated by observation but is the result of a function. However, the feature is easy to copy, for example, by key tracking, if no other properties are used except for the rhythm. Stronger adversaries, like the Russian Impersonation-as-a-Service (IMPaaS) platform~\cite{10.1145/3372297.3417892} or described by Mayrhofer and Sigg~\cite{10.1145/3477601}, could probably record other properties as well. In addition, they could use protocol weaknesses and potential biases in keystroke dynamic patterns. These depend on the smartphone and KDSs. With a bigger distribution, the KDSs are more mainstreamed, making it easier for adversaries. Therefore, the method of keystroke dynamics is not secure enough to protect from these adversaries, especially since the user's smartphone is typically less secure.

\section{\uppercase{Conclusion and Outlook}}
\label{sec:conclusion}

With the increasing demand for stronger authentication methods, biometric authentication systems are being improved and implemented. Due to the growing number of sensors and diverse set of user-driven features, keystroke dynamics is an on-the-go authentication method, which does not hinder the user. But is it good enough? In order to analyze this, we first establish several requirements leading to an evaluation scheme. This evaluation scheme was applied to compare three selected approaches. Based on the comparison, keystroke dynamics and the gathered requirements are discussed. While the pure typing rhythm is too inaccurate, the incorporation of other factors such as pressure strength and size helps to improve the method. In the next step, we plan to extend the current evaluation and analyze and compare the security of keystroke dynamics with traditional authentication methods. A real-world user study helps to evaluate data acquisition under different environmental conditions. This may lead to a common dataset, improving the comparison of different proposed approaches. Last but not least, we evaluate the addition of emojis and other symbols.

\bibliographystyle{apalike}
{\small
\bibliography{keystroke-dynamics}}

\end{document}